\documentclass[prl,amsmath,amssymb,superscriptaddress,twocolumn]{revtex4-1}
\usepackage{graphicx}
\usepackage{amsmath}
\usepackage{amsfonts}
\usepackage{amssymb}
\usepackage{mathrsfs}
\usepackage{bm}
\usepackage{epstopdf}

\newcommand{\be}{\begin{eqnarray}}
\newcommand{\ee}{\end{eqnarray}}

\newcommand{\br}{{\bf r}}

\newcommand{\bk}{{\bf k}}

\newcommand{\bK}{{\bf K}}

\DeclareMathAlphabet{\mathcalligra}{T1}{calligra}{m}{n}
\DeclareMathAlphabet{\mathpzc}{OT1}{pzc}{m}{it} 
\begin{document}
\title{Superconductivity on the brink of spin-charge order in doped honeycomb bilayer}
\author{Oskar  Vafek}
\affiliation{National High Magnetic Field Laboratory and Department
of Physics,\\ Florida State University, Tallahasse, Florida 32306,
USA}
\author{James M. Murray}
\affiliation{National High Magnetic Field Laboratory and Department
of Physics,\\ Florida State University, Tallahasse, Florida 32306,
USA}
\affiliation{Department of Physics and Astronomy,\\ Johns Hopkins University, Baltimore, MD 21218, USA}
\author{Vladimir Cvetkovic}
\affiliation{National High Magnetic Field Laboratory and Department
of Physics,\\ Florida State University, Tallahasse, Florida 32306,
USA}

\begin{abstract}
Using a controlled weak-coupling renormalization group approach, we establish the mechanism of unconventional superconductivity in the vicinity of spin or charge ordered excitonic states for the case of electrons on the Bernal stacked bilayer honeycomb lattice. With one electron per site this system exhibits nearly parabolically touching conduction and valence bands. Such a state is unstable towards a spontaneous symmetry breaking, and repulsive interactions favor excitonic order, such as a charge nematic and/or a layer antiferromagnet. We find that upon adding charge carriers to the system, the excitonic order is suppressed, and unconventional superconductivity appears in its place, before it is replaced by a Fermi liquid. We focus on firmly establishing this phenomenon using the RG formalism within an idealized model with parabolic touching.
\end{abstract}

\date{\today}
\maketitle
Phase diagrams of a number of material classes exhibit unconventional superconductivity in close proximity to N\'eel antiferromagnetism and/or charge ordered states. To date no consensus has emerged regarding the precise mechanism underlying this phenomenon, particularly whether the non-superconducting order is beneficial or detrimental to superconductivity.
While excitonic phases are a natural consequence of repulsive electron-electron interactions, superconductivity is not.
And while it has been advocated that this proximity is not merely a coincidence, and that low energy spin fluctuations or other soft modes from the nearby particle-hole phase tend to enhance superconductivity, there is so far no consensus regarding the precise mechanism by which this occurs \cite{ScalapinoRMP2012,AbanovChubukovFinkelsteinEPL2001,HonerkampEPJ2002,MetlitskiPRB2010a,MetlitskiPRB2010b,EfetovNatPhys2013,Fitzpatrick2013arXiv1307.0004F}.

The bilayer honeycomb lattice in many ways provides an ideal arena in which to explore these questions. To a good approximation, and over a wide energy interval, the conduction and the valence bands touch parabolically at two inequivalent crystal momentum points, $\pm \mathbf{K}$.
The band touching is guaranteed by the time reversal and the crystal symmetries
because the two bands transform as a doublet under the space group operations at $\pm {\bf K}$.
Such a non-interacting state is absolutely unstable to a ground state with a spontaneously broken symmetry once electron-electron interactions are included\cite{VafekYang2010,LemonikPRB2010,VafekPRB2010,NandkishorePRL2010,HonkiMinPRB2008,FanZhangPRL2011,CvetkovicThrockmortonVafekPRB2012,LemonikPRB2012,LangPRL2012}.
Experimental studies on suspended bilayer graphene samples have shown hallmarks of the formation of interaction-driven symmetry breaking (excitonic) phases, with evidence for both gapped\cite{MartinPRL2010,Velasco2012,BaoPNAS2012,FreitagPRL2012,VeliguraPRB2012} and gapless \cite{WeitzScience2010,MartinPRL2010,MayorovvScience2011} behavior. The electron interactions in this system appear to be strong enough to lead to spontaneous symmetry breaking, yet small enough to allow for the use of weakly-coupled theoretical approaches---as evidenced by the small energy scales ($\sim$ meV) up to which the signatures of the ordering behavior appears experimentally.

The central question which we address theoretically here is what happens when additional charge carriers are introduced and the very clean system is driven away from the neutrality point. We find that, as the high energy modes are progressively eliminated, the initially repulsive interaction between electrons turns attractive in several, but not all, two-electron scattering channels.
Importantly, such `repulsion-turned-attraction' happens in the regime which, at weak coupling, can be accessed under strict theoretical control: the strength of the attraction generated is proportional to the strength of the initial repulsion. For small carrier concentration, $\delta n$, the interaction in the attractive {\it and} repulsive channels grows, eventually {\it preventing} superconductivity from occurring, and leading to a state with spin/charge order instead. Increasing the $\delta n$ leads to a saturation of the strength of the repulsive channels, while the attractive channels continue growing, giving rise to a superconducting ground state. Upon further increase of $\delta n$, the attraction is not generated, and the system remains a Fermi liquid. We focus on firmly establishing this result using the RG formalism within an idealized model with parabolic touching, postponing any detailed analysis of its potential experimental observation in a realistic bilayer graphene.
While some theoretical works have proposed supercondutivity in honeycomb bilayer systems, using either t-J model\cite{VucicevicPRB2012,MilovanovicPRB2012} or RPA\cite{FengPRL2013}, unlike ours, such approach is uncontrolled and cannot be used to establish the effect.
\begin{table}[t]
   \begin{tabular}{ | l || l | c || l | l |}
     \hline
      $1_4$ & $\Gamma_1$ & $g_{A1_g}$ & $\Gamma^{(s)}_{7}$ & $\tilde g_{E_{\bf K}}$ \\  \hline
     $\tau_3\sigma_3$ & $\Gamma_2$ & $g_{A2_g}$ & $\Gamma^{(s)}_{8}$ & $\tilde g_{E_{\bf K}}$ \\ \hline
     $1\sigma_1$, $\tau_3\sigma_2$ & $\Gamma_3$,$\Gamma_{10}$ & $g_{E_g}$  & $\Gamma^{(s)}_{5}$, $\Gamma^{(t)}_{15}$ & $\tilde g_{A1_{\bf K}}$, $\tilde g_{A2_{\bf K}}$\\ \hline
     $\tau_31$  & $\Gamma_4$ & $g_{A1_u}$ & $\Gamma^{(s)}_{9}$ & $\tilde g_{E_{\bf K}}$ \\ \hline
     $1\sigma_3$ & $\Gamma_5$ & $g_{A2_u}$ & $\Gamma^{(s)}_{10}$ & $\tilde g_{E_{\bf K}}$ \\ \hline
     $\tau_3\sigma_1$, $-1\sigma_2$ & $\Gamma_6$,$\Gamma_{11}$ & $g_{E_u}$ &$\Gamma^{(s)}_{6}$, $\Gamma^{(t)}_{16}$ & $\tilde g_{A1_{\bf K}}$, $\tilde g_{A2_{\bf K}}$\\ \hline
     $\tau_1\sigma_1$; $\tau_2\sigma_1$ & $\Gamma_7$; $\Gamma_{12}$ & $g_{A1_{\bf K}}$ & $\Gamma^{(s)}_{2}$, $\Gamma^{(t)}_{13}$ &$\tilde g_{E_g}$, $\tilde g_{E_u}$ \\ \hline
     $\tau_1\sigma_2$; $\tau_2\sigma_2$ & $\Gamma_8$;$\Gamma_{13}$ & $g_{A2_{\bf K}}$ & $\Gamma^{(t)}_{14}$, $\Gamma^{(s)}_{3}$ & $\tilde g_{E_u}$, $\tilde g_{E_g}$ \\ \hline
     $\tau_11$,$-\tau_2\sigma_3$& $\Gamma_9$,$\Gamma_{14}$ & $g_{E_{\bf K}}$ & $\Gamma^{(s)}_1$, $\Gamma^{(t)}_{11}$ &
     $\tilde g_{A1_g}$, $\tilde g_{A2_g}$ \\
     $-\tau_21$,$-\tau_1\sigma_3$ &$\Gamma_{15}$,$\Gamma_{16}$ & $g_{E_{\bf K}}$ & $\Gamma^{(t)}_{12}$, $\Gamma^{(s)}_{4}$ &
     $\tilde g_{A1_u}$, $\tilde g_{A2_u}$\\ \hline

   \end{tabular}
\label{tab:matrices}
\caption{Each $\Gamma_j$ in the 2$^{nd}$ column corresponds to one of the 4$\times$4 matrices in the 1$^{st}$ column. Couplings $g_j$ are in the 3$^{rd}$ column. Each representation has the same coupling, e.g. $g_3=g_{10}=g_{E_g}$. The 4$^{th}$ column: the matrices appearing in the pairing bilinears (\ref{eq:H_pp}), with values given in the 1$^{st}$ column. The associated $\tilde g_j$'s are in the last column.
}
\end{table}

We build on the method developed in the Refs.\cite{VafekYang2010,VafekPRB2010,CvetkovicThrockmortonVafekPRB2012} and organize the low energy electronic modes in the vicinity of each valley $\pm {\bf K}$ into a spinor-like object $\psi_{{\bf k},\sigma}=\left(c_{\bf K+k,\sigma},d_{\bf K+k,\sigma},c_{-\bf K+k,\sigma},d_{-\bf K+k,\sigma}\right)^T$, where spin $\sigma=\uparrow\mbox{or} \downarrow$. The annihilation operators $c$ and $d$ correspond to the low-energy electrons in the bottom and top layer, respectively.
The corresponding Hamiltonian operator is
\be
\label{eq:H_ph}
H &=& \sum_{|\mathbf{k}|<\Lambda}\sum_{\sigma=\uparrow,\downarrow}
	\psi^\dagger_{\mathbf{k},\sigma} \mathcal{H}_\mathbf{k} \psi_{\mathbf{k},\sigma}+ \frac{2\pi}{m^*}\sum_{j=1}^{16} g_j \int d^2 {\bf r}
	\rho^2_j(\br),
\ee
where $\rho_j(\br)=\sum_{\sigma=\uparrow,\downarrow} \psi_{\sigma}^\dagger ({\bf r}) \Gamma_j \psi_{\sigma} ({\bf r})$
and $\psi_{\sigma}({\bf r})=\frac{1}{L}\sum_{|\mathbf{k}|<\Lambda}e^{i\mathbf{k}\cdot \mathbf{r}} \psi_{\mathbf{k},\sigma}$, $L$ is the linear system size and \be
\mathcal{H}_\mathbf{k} =& \frac{k_x^2 - k_y^2}{2m^*} 1\sigma_1
	+ \frac{k_x k_y}{m^*} \tau_3\sigma_2 + v_3 k_x \tau_3\sigma_1 - v_3 k_y 1\sigma_2.
\ee
The Pauli matrices $\tau_i$ and $\sigma_i$ operate in valley and layer spaces, respectively.
There are 16 independent 4$\times$4 matrices denoted here by $\Gamma_j$.
Of the 16 dimensionless couplings $g_j$ only the first 9 are independent, each corresponding to an irreducible representations of the lattice space group:
${\bf D}_{3d}$ at $\Gamma=(0,0)$, and ${\bf D}_3$ at ${\bf K}=\left(\frac{4\pi}{3\sqrt{3}a},0\right)$.
For bilayer graphene, $m^* \approx 0.029m_e$ \cite{MayorovvScience2011,Velasco2012,WeitzScience2010}, while $v_3$, which at very low energies distorts the parabolic spectrum into four Dirac cones near each of the $\pm \mathbf{K}$ points, is\cite{MayorovvScience2011} $v_3 \approx 1.41\times 10^5$ m/s. The upper cutoff energy scale $\Omega_{\Lambda}=\Lambda^2 / 2m^* \sim 0.2$ eV.

We rewrite the equilibrium partition function $Z=\mbox{Tr}\left(e^{-\beta\left(H-\mu N\right)}\right)$ in terms of the usual coherent state Grassman path integral
$Z=\int\mathcal{D}(\psi^*,\psi)e^{-\mathcal{S}}$, where $\beta=1/k_BT$. The action $\mathcal{S}=\int_0^{\beta}\left(\sum_{\bk,\sigma}\psi_{\bk,\sigma}^{\dagger}(\tau)\left(\frac{\partial}{\partial \tau}-\mu\right)\psi_{\bk,\sigma}(\tau)+H(\tau)\right)$ and $H(\tau)$ is obtained by replacing the operators $\psi_{\bk,\sigma}$ with Grassman valued fields $\psi_{\bk,\sigma}(\tau)=\frac{1}{\beta}\sum_{n=-\infty}^{\infty}e^{-i\omega_n\tau}\psi_{\bk,\sigma,n}$ with Matsubara frequencies $\omega_n=(2n+1)\pi k_BT$.
The Wilsonian RG procedure begins by integrating out all $\psi_{\bk,\sigma,n}$'s within a thin shell of momenta $(1-d\ell)\Lambda < |\bk| < \Lambda$, but for all $n$ and $\sigma$. Afterwards, we rescale the momenta, the fields, temperature, the chemical potential $\mu$, and $v_3$ in such a way that the cutoff goes back to $\Lambda$ and the terms in $\mathcal{H}_{\bk}$, which are quadratic in $\bk$, remain invariant. According to this `tree-level' rescaling, one finds that the coupling constants $g_j$ remain unchanged while $T_\ell = T e^{2\ell}$, $\mu_\ell = \mu e^{2\ell}$, and $v_{3\ell} = v_3 e^{\ell}$.
For any finite range interactions, therefore, $H$ is of the most general form respecting particle-hole symmetry, in that omitted interaction terms containing derivatives or product of more than four fermion fields quickly renormalize to zero.

At the next order in $g_j$, the RG flows of $T$ and $v_3$ are unaffected by the interaction correction.
The modification appears in the flows of $\mu_{\ell}$ and $g_j$'s. We will now set $T=v_3=0$. Then the flow of $\mu_{\ell}$ does not change and
\begin{equation}
\label{eq:g_flow}
\frac{dg_i(\ell)}{d\ell} = \sum_{j,k=1}^9 \mathcal{A}_{ijk}(\mu_\ell) g_j(\ell) g_k(\ell).
\end{equation}
These flow equations are valid as long as $g_i \ll 1$.

\begin{figure}
\hspace{-.5cm} \includegraphics[width=0.5\textwidth]{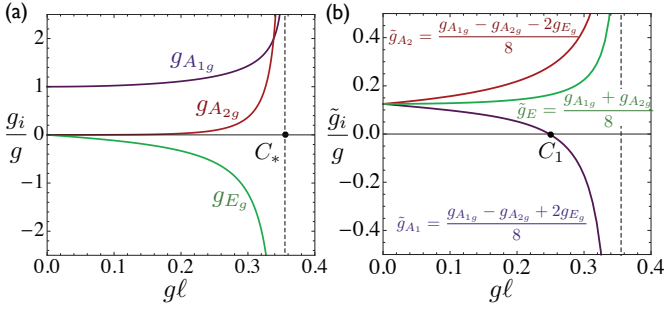}
 \caption{(a) The flow of interaction couplings $g_j$ defined in the Eq.\ref{eq:H_ph}, when initially $g=g_{A1g}(0)>0$, while all other couplings vanish. Under such conditions, only $g_{A2g}$ and $g_{Eg}$ get generated, while all other couplings vanish. Here $\mu=0$.
  (b) The corresponding flow of the Cooper pair couplings $\tilde g_j$ defined in the Eq.\ref{eq:H_pp}. All nine $\tilde g_j$'s are finite as can be seen from the Fierz matrix \ref{eq:Fierz}; their values are independent of the $g$,$u$, and $\bK$ label i.e. $\tilde g_{A1g}=\tilde g_{A1u}=\tilde g_{A1\bK}=\tilde g_{A1}$, etc. Note that if $g$ is small then the attractive interactions are generated in the regime where the weak coupling RG is fully justified.
  \label{fig:fwd scatt g}}
\end{figure}

At the neutrality point $\mu_{\ell}=0$ and the functions $\mathcal{A}_{ijk}$ reduce to real constants. The equation (\ref{eq:g_flow}) is left invariant upon
simultaneous rescaling of $g_j\rightarrow b g_j$ and $\ell \rightarrow \ell/b$. Therefore, any solution of
the Eqs. (\ref{eq:g_flow}) has the form
\begin{eqnarray}\label{eq: scaling form of g}
g_{i}(\ell,\{g_j(0)\})=g \Phi_i\left(g\ell,\{g_j(0)/g\}\right),
\end{eqnarray}
where $g=\sqrt{\sum_{j} g^2_{j}(0)}$ and $\Phi_i$'s are
functions which can be determined numerically, which typically diverge at a finite value of $\ell=\ell_*$. Key insight can be gained by {\it exactly} recasting the interaction term in the action as a pairing interaction $\mathcal{S}_\mathrm{int}=\frac{2\pi}{m^*}\int_0^{\beta}d\tau\int d^2 {\bf r}\mathcal{L}_\mathrm{int}$ and
\begin{equation}
\label{eq:H_pp}
\mathcal{L}_\mathrm{int}=
\sum_{j=1}^{10}\tilde g_j  \mbox{S}^\dagger_j({\bf r},\tau)\mbox{S}_j({\bf r},\tau)+\sum_{j=11}^{16} \tilde g_j \vec{\mbox{T}}^\dagger_j({\bf r},\tau)\cdot \vec{\mbox{T}}_j({\bf r},\tau)
\end{equation}
where the spin singlet and the spin triplet Cooper pair terms are
\begin{eqnarray}
\mbox{S}_j({\bf r},\tau)&=& \sum_{\alpha,\beta=\uparrow,\downarrow}
\psi_{\alpha}^T ({\bf r},\tau) \Gamma^{(s)}_j\left(i\sigma_2\right)_{{\alpha\beta}} \psi_{\beta} ({\bf r},\tau),\\
\vec{\mbox{T}}_j({\bf r},\tau)&=& \sum_{\alpha,\beta=\uparrow,\downarrow}
\psi_{\alpha}^T ({\bf r},\tau) \Gamma^{(t)}_j\left(i\sigma_2\vec{\sigma}\right)_{{\alpha\beta}} \psi_{\beta} ({\bf r},\tau).
\end{eqnarray}
The 9 independent pair interactions, $\tilde g_j$, can be written as a linear combination of $g_j$'s using Fierz identities
\begin{eqnarray}\label{eq:Fierz}
\tilde g_{R_p}=\sum_{R'=A1,A2,E} \mathcal{F}_{RR'}\sum_{p'=g,u,\bK}\mathcal{F}_{pp'} g_{{R'}_{p'}}
\end{eqnarray}
where $\mathcal{F}=\left(\begin{array}{ccc}
1 & -1 & 2\\
1 & -1 & -2\\
1 & 1 & 0
\end{array}
\right)$.
For generic repulsive interactions {\it all} $\tilde g_j$'s are initially repulsive and not obviously conducive to Cooper pairing. Nevertheless, under RG attraction is generated: there is a scale $\ell_1$ where $\tilde g_i(\ell_1)=0$ for some $i$'s, and continues negative for $\ell_1<\ell<\ell_*$. An example of this can be seen in the Fig.\ref{fig:fwd scatt g} where $g\equiv g_{A1_g}(0)>0$ otherwise $g_{j}(0)=0$ (we refer to this as the forward scattering limit). Flow equations for this model at $\mu=0$ can be found in Ref.\onlinecite{VafekYang2010}.
Due to the scaling form discussed above, $\ell_1=C_1/g$, and similarly $\ell_*=C_*/g$, where $C_*>C_1>0$. At $\ell_1$ the couplings therefore attain values $g_i(\ell_1)=g \Phi_i\left(C_1,\{g_j(0)/g\}\right)$.
Since $\Phi_i\left(C_1,\{g_j(0)/g\}\right)$ are finite numbers, independent of $g$, we arrive at an important conclusion that if $g$ is small then so is $g_i(\ell_1)$; attraction is therefore generated in the regime when the flow equations (\ref{eq:g_flow}) are valid.
As long as $\mu_{\ell_1}\ll \Omega_{\Lambda}$, a finite $\mu$ does not change the above conclusion, because it has essentially no effect on the flow equations up to, and near, $\ell_1$.

\begin{figure}
 \includegraphics[width=0.5\textwidth]{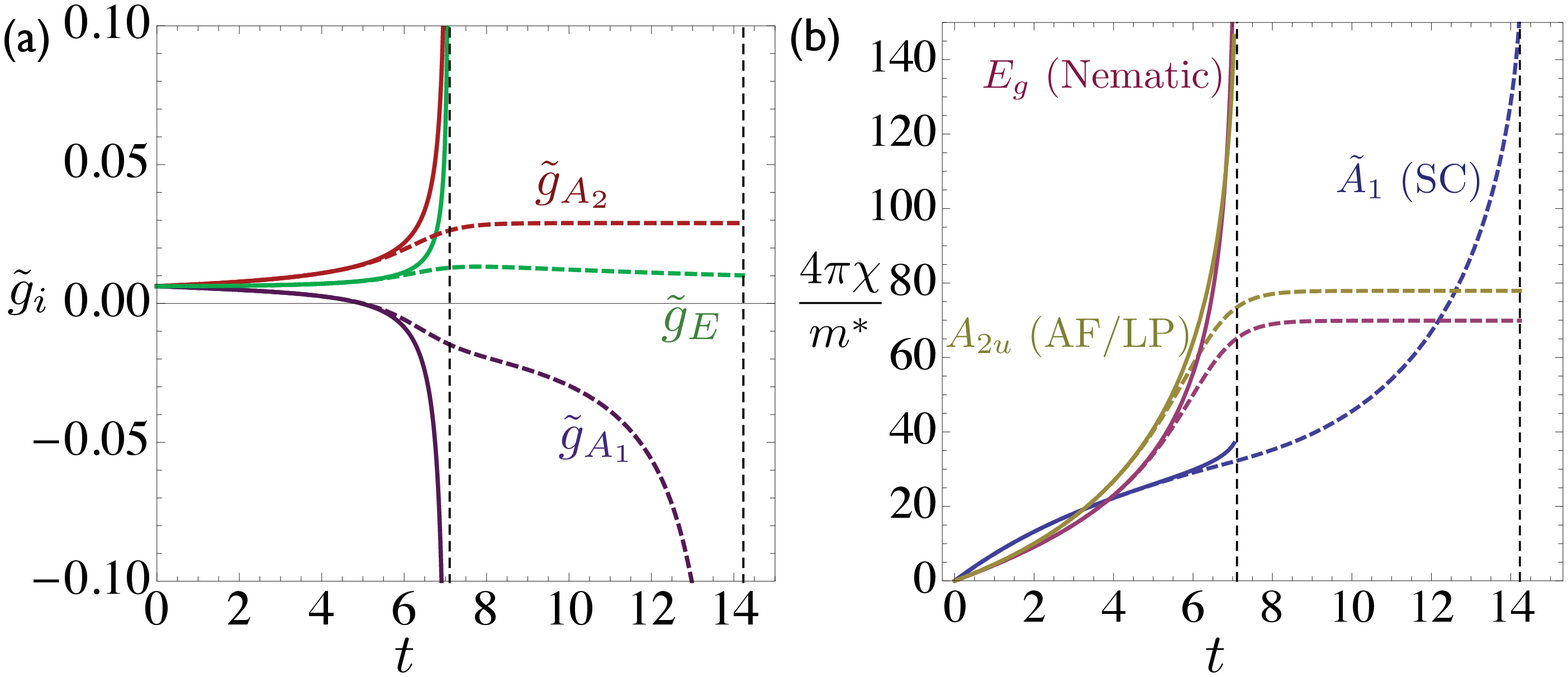}
 \caption{RG flows with $g_{A1g}(0)=0.05$ vs. $t$ (Eq.\ref{eq: t def}). For the solid curves $\mu=10^{-9}\Omega_{\Lambda}$, for the dashed curves $\mu=3\times 10^{-6}\Omega_{\Lambda}$. The former is in the regime $\mu\ll\Omega_{\Lambda} e^{-2C_*/g}$ leading to an excitonic order, the latter in $\Omega_{\Lambda}e^{-2C_*/g} \ll \mu\ll \Omega_{\Lambda}e^{-2C_1/g}$, leading to superconductivity. (a) $\tilde g$ in Eq.\ref{eq:H_pp}. (b) Susceptibilities in excitonic and superconducting channels.
  \label{fig:fwd scatt g vs t}}
\end{figure}

However, such attractive pair interactions do not {\it necessarily} lead to superconductivity. As shown in the Fig.\ref{fig:fwd scatt g}, if $\mu_{\ell_*}\ll \Omega_{\Lambda}$ the growth of the attractive $\tilde g$'s is accompanied by the growth of the repulsive $\tilde g$'s, disfavoring superconductivity and favoring an excitonic state. In order to demonstrate this, we introduce infinitesimal symmetry breaking source terms into the starting Hamiltonian, $H\rightarrow H+\sum_{\bk}\left(\sum^{16}_{j=1}\delta H^{(j)}_1+\sum^{10}_{j=1}{\Delta}^{pp}_{j}\delta H^{(j)}_{2s}+\sum^{16}_{j=11}\vec{\Delta}^{pp}_{j}\cdot\delta H^{(j)}_{2t}\right)$, where
$\delta H^{(j)}_1=  \psi^\dagger_{\bk,\alpha}\left(\Delta_j^{ph}\Gamma_j\delta_{\alpha\beta}+\vec{\Delta}_j^{ph}\cdot\Gamma_j\vec{\sigma}_{\alpha\beta}\right) \psi_{\bk,\beta}$, and
$\delta H^{(j)}_{2(s,t)} =\frac{1}{2}
\psi^T_{\bk,\alpha} \Gamma^{(s,t)}_j\left(i\sigma_2\left(1,\vec{\sigma}\right)\right)_{\alpha\beta}\psi_{-\bk,\beta}+ h.c.$
Using our RG procedure we find the dependence of the Helmholtz free energy, $\delta f$,  on $\Delta_j$'s and $\ell$.
This is then used to compute the susceptibility,
$\chi_{ij}=-\frac{\partial^2 \delta f}{\partial \Delta^*_i \partial \Delta_j}$,
associated with excitonic or superconducting ordering tendencies shown in the Fig.\ref{fig:fwd scatt g vs t}.
We see that in the regime $\mu\ll \Omega_{\Lambda} e^{-2C_*/g}$, despite generating the attractive interactions at $\ell_1$, the susceptibility in the excitonic channels grows above the superconducting ones. With pure forward scattering, the dominant instability appears to be the charge nematic\cite{VafekYang2010,LemonikPRB2010,LemonikPRB2012,CvetkovicThrockmortonVafekPRB2012}.

On the other hand, if $\mu_{\ell_*}\gg \Omega_{\Lambda} \gg \mu_{\ell_1}$, then, once generated, the attractive interactions continue growing while the repulsive ones do not. The Fermi surface is reached after $\ell_1$ but before $\ell_*$, and the ground state is a superconductor. Substituting for $\ell_1$ and $\ell_*$, the condition translates into $\Omega_{\Lambda}e^{-2C_*/g} \ll \mu\ll \Omega_{\Lambda}e^{-2C_1/g}$. Because $C_1<C_*$, this can always be satisfied for sufficiently small $g$.
Indeed, the flow equations for $\tilde g_j$'s take the form
\begin{eqnarray}
\frac{d\tilde g_i}{d\ell}=\frac{-a_i}{\left(1-\frac{\mu}{\Omega_{\Lambda}} e^{2\ell}\right)}\tilde g^2_i + \sum_{j,k} \tilde A_{ijk}\left(\frac{\mu}{\Omega_{\Lambda}} e^{2\ell}\right)\tilde g_j \tilde g_k
\end{eqnarray}
where $a_i\geq 0$ and the functions $\tilde A_{ijk}$ are non-singular at $\ell_{FS}=\frac{1}{2}\ln \frac{\Omega_{\Lambda}}{\mu}$.
This result is general.
Letting
\begin{eqnarray}\label{eq: t def}
t=\frac{1}{2}\ln \left(\frac{\Omega_{\Lambda}-\mu}{\Omega_{\Lambda}e^{-2\ell}-\mu}\right),
\end{eqnarray}
which vanishes at $\ell=0$ and grows without bound as $\ell\rightarrow\ell_{FS}$
the flow equations take the form
\begin{eqnarray}
\frac{d\tilde g_i}{dt}=-a_i \tilde g^2_i+\frac{\sum_{j,k}\tilde A_{ijk}\left(\frac{\mu}{\Omega_{\Lambda}} e^{2\ell(t)}\right)\tilde g_j \tilde g_k}{1+\frac{\mu}{\Omega_{\Lambda}-\mu}e^{2t}}.
\end{eqnarray}
The second term is just as important as the first for small $t$. However, as $\ell$ approaches $\ell_{FS}$, $t$ grows and the second term is exponentially suppressed. Because $\ell_{FS}<\ell_*$, all the couplings are of order $g$ at the beginning of the regime marked by $t_0$ where the second term may be neglected. The solution of the resulting equation is $\tilde g_i(t)=\tilde g_i(t_0)/(1+a_i\tilde g_i(t_0)(t-t_0))$. The negative couplings thus grow and become of $\mathcal{O}(1)$ when $t-t_0$ is $\mathcal{O}\left(1/\tilde g_i(t_0)\right)\sim \mathcal{O}\left(1/g\right)$. Since $\mu e^{2t_0}/\Omega_{\Lambda}$ is of $\mathcal{O}(1)$, the second term is indeed exponentially small and may be neglected. The repulsive couplings are therefore small when the attractive couplings become of $\mathcal{O}(1)$. The only $\chi$'s which grow (with mean-field exponents) are therefore in the attractive Cooper channels, see Fig.\ref{fig:fwd scatt g vs t}.
In the third regime, $\mu_{\ell_1}\gg \Omega_{\Lambda}$, and the attraction is not generated; the system is a Fermi liquid.

Generalizing to the case of microscopic density-density interactions, the only additional nonzero bare couplings besides $g_{A_{1g}}$ are $g_{A_{2u}}$ and $g_{E_\mathbf{K}}$, which correspond to interlayer scattering and backscattering, respectively \cite{throckmorton12}. Increasing the latter two relative to $g_{A_{1g}}$ corresponds to decreasing the spatial range of the interaction, with the Hubbard limit of on-site interaction corresponding to $g_{A_{2u}} = 2g_{E_\mathbf{K}} = g_{A_{1g}}$. In order to interpolate between the two limits, we let $g_{A_{2u}}=2 g_{E_\mathbf{K}} = \lambda g_{A1g}$, so that $\lambda = 0$ and $\lambda = 1$ correspond to the forward-scattering and Hubbard cases, respectively.
\begin{figure}[t]
\centering
\hspace{-0.5cm}\includegraphics[width=0.5\textwidth]{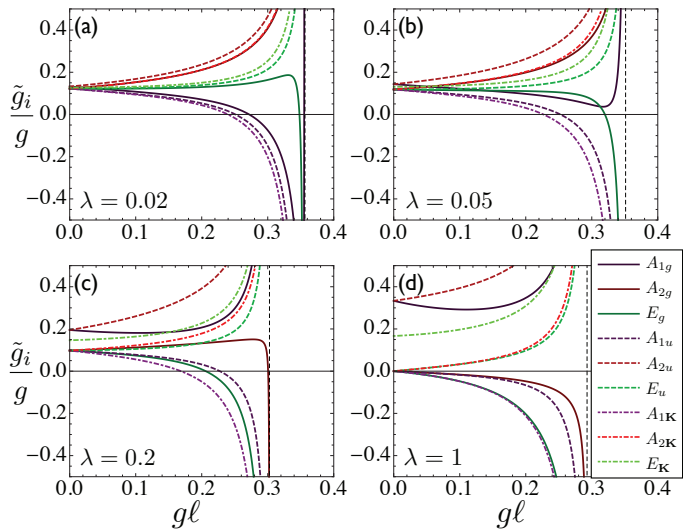}
\caption{Coupling flows for various values of bare couplings, with $\mu = 0$. $\lambda=0$ corresponds to the forward-scattering limit, in which only $g_{A_{1g}}$ is nonzero, while $\lambda = 1$ corresponds to on-site Hubbard interaction.
\label{fig:flows_4panel}}
\end{figure}
For small $\lambda$, shown in Fig.\ref{fig:flows_4panel}(a), the $\mu=0$ flows are similar to those in Fig.\ref{fig:fwd scatt g}, with the three $\tilde g$'s corresponding to $A_1$, $A_2$, and $E$ each remaining nearly degenerate over most of the range. The first three $\tilde g$'s to turn negative are again the $A_1$, although they no longer cross at the same point, with $\tilde g_{A_{1\mathbf{K}}}$ changing sign first. Near the flow singularity at $g\ell = C_* \approx 0.356$, the $d$-wave ($E_g$) and $s$-wave ($A_{1g}$) couplings turn negative and positive, respectively. As $\lambda$ is increased further, $\tilde g_{E_g}$ becomes negative at progressively smaller values of $\ell$, while $\tilde g_{A_{1g}}$ remains repulsive over the entire range of flows for $\lambda \gtrsim 0.05$, indicating that even a small amount of backscattering suppresses tendency toward $s$-wave pairing. The $\tilde g_{A_{2g}}$ coupling also turns attractive for sufficiently large $\lambda$.

The case of Hubbard interaction is shown in Fig.\ref{fig:flows_4panel}(d).
At $\mu=0$ the dominant instability appears in the layer anti-ferromagnetic channel\cite{VafekPRB2010,CvetkovicThrockmortonVafekPRB2012,LangPRL2012}.
It follows from Eq.\ \ref{eq:Fierz} that many of the bare couplings $\tilde g_i$ vanish, so that some will become negative already at infinitesimal $\ell$. Thus, according to the arguments in the previous section, at $T=0$ one no longer obtains a Fermi liquid at any $\mu$. The Hubbard model is special in that it exhibits a superconducting instability even for large values of $\mu$, somewhat similar to the square lattice case\cite{ZanchiSchulzPRB1996,HlubinaPRB1999,MrazHlubinaPRB2003,RaghuKivelsonScalapinoPRB2010}. From the flows in Figure \ref{fig:flows_4panel}(d), we see that the $\tilde g_{A_{1\mathbf{K}}}$, corresponding to a pair density wave (PDW), and $\tilde g_{E_g}$, corresponding to $d$-wave, superconducting channels are the most attractive. We analyzed the $\chi$'s, and solved the self-consistent mean field equations with $\tilde g$'s and $\mu$ determined by terminating the RG flow at $t$ where the $\tilde g$ are decoupled. We find that PDW emerges as the leading instability.
This corresponds to a unidirectional $2\bK$-modulation of the real singlet pairing amplitude, with a fully gapped spectrum.
Since such state is dependent upon the circular symmetry of each Fermi pocket, however, we expect that the $d\pm id$-wave state will be favored for a more physically realistic model, in which this symmetry is destroyed by further neighbor hopping, or equivalently $v_3\neq 0$. \cite{MurrayUnpublished}

\begin{figure}
 \includegraphics[width=0.48\textwidth]{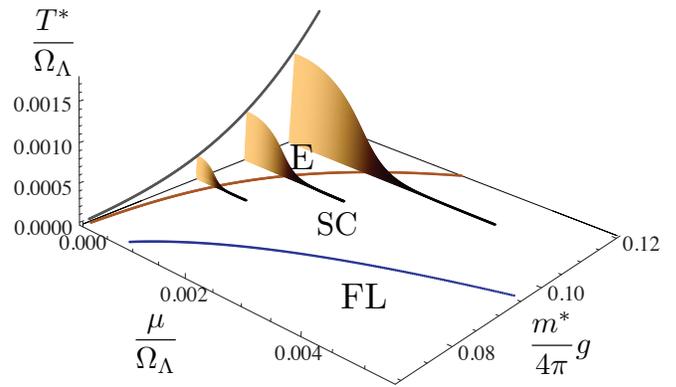}
 \caption{Temperature $T^*$ associated with ordering vs. $\mu$ and $g$ for the forward-scattering limit. The crossover lines shown correspond to asymptotic bounds for the phase boundaries between excitonic (E), superconducting (SC), and Fermi liquid (FL) states.
  \label{fig:phase diag}}
\end{figure}

We estimate the crossover temperature $T^*$ associated with ordering from the value of $\ell=\ell_\mathcal{O}$ where couplings become of $\mathcal{O}(1)$ to be the solution of $2\cosh\left(\frac{\mu}{T^*}\right)\exp\left(-\Omega_{\Lambda} e^{-2\ell_\mathcal{O}}/T^*\right)\approx 1$. This comes from estimating the temperature at which the thermal factors appearing in the finite $T$ flow equations at the scale $\ell_{\mathcal{O}}$ deviate appreciably from their low temperature asymptotic form\cite{MurrayUnpublished}. The dependence of $T^*$ on $\mu$ and $g$ for the forward scattering limit is shown in Fig.\ref{fig:phase diag}.

Finally, given that the paired states found here are unconventional, they are sensitive to disorder. This may be the reason for currently no reports of superconductivity in materials with bilayer honeycomb structure. Nevertheless, given that the doping can be controlled electrostatically, and that some of the high purity samples have started showing signatures of excitonic ordering, further increase in sample purity may be a promising avenue towards achieving superconductivity.

This work was supported by the NSF CAREER award under
Grant No. DMR-0955561 (OV), NSF Cooperative Agreement No.
DMR-0654118, and the State of Florida (OV,JM,CV), as well as by ICAM-I2CAM (NSF grant DMR-0844115) and by DoE, Office of Basic Energy Sciences, Division of Materials Sciences and Engineering under Award DE-FG02-08ER46544 (JM).

\bibliography{blgsc}

\newpage
\begin{appendix}
\begin{widetext}
\section{Fierz Identities}
Any $2\times2$ matrix $A$ can be expanded in terms of the unit matrix, $\sigma^{(0)}$, and the three Pauli matrices as
$A_{\mu\nu}=\frac{1}{2}\sum_{j=0}^3A_{\alpha\beta}\sigma^{(j)}_{\beta\alpha}\sigma^{(j)}_{\mu\nu}$, where the repeated indices are assumed summed over.
Since each element of $A$ is arbitrary, we can take derivative with respect to each independently. Using this procedure we find
that $\delta_{\mu\mu'}\delta_{\nu\nu'}=\frac{1}{2}\sum_{j=0}^3\sigma^{(j)}_{\mu\nu}\sigma^{(j)}_{\nu'\mu'}$.

Similarly, any $4\times 4$ matrix $M$ can be expanded in terms of the sixteen matrices $\Gamma_j$ as
$M_{\mu\nu}=\frac{1}{4}\sum_{j=1}^{16} M_{\alpha\beta}\left(\Gamma_{j}\right)_{\beta\alpha}\left(\Gamma_j\right)_{\mu\nu}$.
We also have $\delta_{\mu\mu'}\delta_{\nu\nu'}=\frac{1}{4}\sum_{j=1}^{16} \left(\Gamma_{j}\right)_{\mu\nu}\left(\Gamma_{j}\right)_{\nu'\mu'}$.

Using these identities we find
\begin{eqnarray}
M_{\alpha\beta}M_{\gamma\delta}=\frac{1}{16}\sum_{k=1}^{16}\sum_{k=1}^{16}\left(\Gamma_{j}\right)_{\alpha\gamma}\left(\Gamma_{k}\right)_{\delta\beta}
\mbox{Tr}\left[\Gamma_j M \Gamma _k M^{T}\right].
\end{eqnarray}
This implies the following relation between the Grassman valued expressions
\begin{eqnarray}
\psi^{\dagger}_{\sigma}M\psi_{\sigma}\psi^{\dagger}_{\sigma'}M\psi_{\sigma'}=
\frac{1}{32}\sum_{i=0}^3\sum_{j=1}^{16}\sum_{k=1}^{16}\mbox{Tr}\left[\Gamma_j M \Gamma_k M^T\right]
\left(\psi^{\dagger}_{\alpha}\Gamma_j\sigma^{(i)}_{\alpha\beta}\psi^*_{\beta}\right)\left(\psi^{T}_{\alpha'}\Gamma_k\sigma^{(i)}_{\alpha'\beta'}\psi_{\beta'}\right).
\end{eqnarray}
The Grassman nature of the $\psi$ fields implies that the only non-vanishing terms in the triple sum are those which satisfy $\left(\Gamma_{j}\right)_{\alpha\alpha'}\sigma^{(i)}_{\beta\beta'}=-\left(\Gamma_{j}\right)_{\alpha'\alpha}\sigma^{(i)}_{\beta'\beta}$.
Such identities lead to the relation between the nine couplings
\begin{eqnarray}\label{eq:FierzAppendix}
\left(\begin{array}{c}
\tilde g_{A1_g} \\
\tilde g_{A2_g} \\
\tilde g_{E_g} \\
\tilde g_{A1_u} \\
\tilde g_{A2_u} \\
\tilde g_{E_u} \\
\tilde g_{A1_{\bf K}} \\
\tilde g_{A2_{\bf K}} \\
\tilde g_{E_{\bf K}}
\end{array}\right)=\frac{1}{8}
\left(\begin{array}{ccccccccc}
1 & -1 & 2 & -1 & 1 & -2 & 2 & -2 & 4 \\
1 & -1 & -2 & -1 & 1 & 2 & 2 & -2 & -4 \\
1 & 1 & 0 & -1 & -1 & 0 & 2 & 2 & 0 \\
1 & -1 & 2 & -1 & 1 & -2 & -2 & 2 & -4 \\
1 & -1 & -2 & -1 & 1 & 2 & -2 & 2 & 4 \\
1 & 1 & 0 & -1 & -1 & 0 & -2 & -2 & 0 \\
1 & -1 & 2 & 1 & -1 & 2 & 0 & 0 & 0 \\
1 & -1 & -2 & 1 & -1 & -2 & 0 & 0 & 0 \\
1 & 1 & 0 & 1 & 1 & 0 & 0 & 0 & 0
\end{array}
\right)
\left(\begin{array}{c}
g_{A1_g} \\
g_{A2_g} \\
g_{E_g} \\
g_{A1_u} \\
g_{A2_u} \\
g_{E_u} \\
g_{A1_{\bf K}} \\
g_{A2_{\bf K}} \\
g_{E_{\bf K}}
\end{array}\right),
\end{eqnarray}
presented in the Eq.\ref{eq:Fierz} of the text.

\section{Flow equations in the forward scattering limit for $T=v_3=0$}
The Fourier transform of the imaginary time Greens function is
\begin{eqnarray}
G_{\bk}(i\omega)&=&\left(\left(-i\omega+\mu\right)1_4+\frac{k^2_x-k^2_y}{2m^*}1\sigma_1+\frac{k_xk_y}{m^*}\tau_3\sigma_2 \right)^{-1}\nonumber\\
&=&\frac{\left(i\omega-\mu\right)1_4+\frac{k^2_x-k^2_y}{2m^*}1\sigma_1+\frac{k_xk_y}{m^*}\tau_3\sigma_2}{\left(\omega+i\mu\right)^2+\left(\frac{\bk^2}{2m^*}\right)^2}
\end{eqnarray}
where $\omega_n=(2n+1)\pi T$ is the Matsubara frequency.
The following pair of identities is useful in deriving the flow equations
\begin{eqnarray}
&&\int_{\Lambda(1-d\ell)}^{\Lambda}\frac{dkk}{2\pi}\int_0^{2\pi}\frac{d\theta_{\bk}}{2\pi}\int_{-\infty}^{\infty}\frac{d\omega}{2\pi}
G_{\bk}(i\omega)\otimes G_{\bk}(i\omega)=\nonumber\\
&&\frac{m^*}{4\pi}d\ell F_0\left(\frac{\mu}{\Lambda^2/2m^*}\right)
\left(-1_4\otimes 1_4+\frac{1}{2}\left(1\sigma_1\otimes 1\sigma_1+\tau_3\sigma_2\otimes\tau_3\sigma_2\right)\right)\\
&&\int_{\Lambda(1-d\ell)}^{\Lambda}\frac{dkk}{2\pi}\int_0^{2\pi}\frac{d\theta_{\bk}}{2\pi}\int_{-\infty}^{\infty}\frac{d\omega}{2\pi}
G_{\bk}(i\omega)\otimes G_{-\bk}(-i\omega)=\nonumber\\
&&\frac{m^*}{4\pi}d\ell
\left(F_1\left(\frac{\mu}{\Lambda^2/2m^*}\right)1_4\otimes 1_4+\frac{1}{2}F_2\left(\frac{\mu}{\Lambda^2/2m^*}\right)\left(1\sigma_1\otimes 1\sigma_1+\tau_3\sigma_2\otimes\tau_3\sigma_2\right)\right)\end{eqnarray}
where for $x>0$
\begin{eqnarray}\label{eq: Fs}
F_0(x)&=&\frac{1}{2}\mbox{sign}\left(1+x\right)+\frac{1}{2}\mbox{sign}\left(1-x\right),\\
F_1(x)&=&\frac{1}{2x}\left(\frac{1+2x}{1+x}-\frac{1-2x}{|1-x|}\right),\\
F_2(x)&=&-\frac{1}{2x}\left(\frac{1}{1+x}-\frac{1}{|1-x|}\right).
\end{eqnarray}
Note that $F_1(x)=F_2(x)$ for $0\leq x\leq 1$.

In the forward scattering limit only $g_{A1_g}$ is assumed non-zero initially. As shown in the Ref.\onlinecite{VafekYang2010}, there are two other couplings which are generated under RG, i.e. $g_{A2_g}$ and $g_{E_g}$.
The above results lead to the following equations $(N=4)$
\begin{eqnarray}\label{eq: g fwd 1}
\frac{d g_{A1g}}{d\ell}&=&\left(\left(g_{A1g}^2+g_{A2g}^2+2 g_{Eg}^2 \right)(F_0-F_1)-2 g_{A1g} g_{Eg}
   (F_0+F_2)+2
   g_{A2g} g_{Eg} (F_2-F_0)\right),\\
\frac{d g_{A2g}}{d\ell}&=&
 \left(2g_{A1g}g_{A2g} \left(3F_0-F_1\right)+2g_{A1g}g_{Eg} \left(F_2-F_0\right)-4(N-1)
   g_{A2g}^2 F_0\right.\nonumber\\
&& \left.-2g_{A2g} g_{Eg} (5
   F_0+F_2)+2g_{Eg}^2 (F_0+F_1)\right),\label{eq: g fwd 2}\\
\frac{d g_{Eg}}{d\ell}&=&
 \frac{1}{2}
\left(-g_{A1g}^2 (F_0+F_2)+ 2g_{A1g}g_{A2g} (F_2-F_0)+4g_{A1g}g_{Eg}
   (2F_0-F_1)-g_{A2g}^2
   (F_0+F_2)\right.\nonumber\\
&&\left. +4 g_{A2g} g_{Eg} F_1-4
   g_{Eg}^2 [(N+1)F_0+F_2]\right),
   \label{eq: g fwd 3}
\end{eqnarray}
where we omitted the argument of each $F_i$, which is $\hat{\mu}_{\ell}=\frac{\mu}{\Lambda^2/2m^*}e^{2\ell}$.
At $\mu=0$ these equations reduce to those found in Ref.\onlinecite{VafekYang2010}.
Assuming that only $g_{A1_g}(0)\neq 0$, the value of $C_1$ can be found rather precisely using series expansion techniques to be $C_1=0.248498$.
Similarly, the value of $C_*=0.3553$. These values are indicated in the Fig.\ref{fig:fwd scatt g}.
The above equations can be rewritten in terms of $\tilde g_j$'s using Eq.(\ref{eq:Fierz}). We find
\begin{eqnarray}\label{eq: g tilde fwd 1}
\frac{d\tilde g_{A1}}{d\ell}&=&-8 {\tilde g_{A1}}^2 (F_1+F_2)\nonumber\\
&-&8\left[-{\tilde g_{A1}}\left(({\tilde g_{A2}}-{\tilde g_{E}}) N+{\tilde g_{A2}}+3{\tilde g_{E}}\right)+{\tilde g_{E}}(({\tilde g_{A2}}-{\tilde g_{E}})N+{\tilde g_{A2}}+3 {\tilde g_{E}})\right]F_0\\
\frac{d\tilde g_{A2}}{d\ell}&=&-8\tilde g^2_{A2g}(F_1-F_2)\nonumber\\
&+&4 \left[{\tilde g_{A1}}^2 (N-1)+{\tilde g_{A2}}^2 (N+3)+2 {\tilde g_{E}} \left(\tilde g_{E}-\tilde g_{A1}-\tilde g_{A2} \right)(N-1)\right]F_0
\label{eq: g tilde fwd 2}\\
\frac{d\tilde g_{E}}{d\ell}&=&-8 {\tilde g_{E}}^2 F_1\nonumber\\
&-&2 \left[N
   ({\tilde g_{A1}}+{\tilde g_{A2}}-2 {\tilde g_{E}})^2-3 {\tilde g_{A1}}^2+2 {\tilde g_{A1}} {\tilde g_{A2}}
   +12 {\tilde g_{A1}} {\tilde g_{E}}+{\tilde g_{A2}}^2-4 {\tilde g_{A2}} {\tilde g_{E}}-12 {\tilde g_{E}}^2
   \right]F_0.\nonumber
   \label{eq: g tilde fwd 3}\\
\end{eqnarray}
where, again, we omitted the argument of each $F_i$, which is $\hat{\mu}_{\ell}=\frac{\mu}{\Lambda^2/2m^*}e^{2\ell}$, as well as the specification of the $g$, $u$ or $\bK$ label on $\tilde g$, since in this case the result is independent of it.
The information contained in Eqs.(\ref{eq: g fwd 1}-\ref{eq: g fwd 3}) and in Eqs.(\ref{eq: g tilde fwd 1}-\ref{eq: g tilde fwd 3}) is identical. This can be seen by performing the susceptibility analysis.
However, the latter form is more transparent in the regime $\frac{\Lambda^2}{2m^*}e^{-2C_*/g}\ll \mu \ll \frac{\Lambda^2}{2m^*}e^{-2C_1/g}$ due to the manifest decoupling of the equations in the vicinity of the Fermi surface, i.e. when $\ell\approx \sqrt{\frac{\Lambda^2/2m^*}{\mu}}$, as discussed in the main text.

\section{Susceptibility from the flow equations for $T=v_3=0$.}
\label{appendix susceptibility}
The flow equations for the particle-hole source terms are
\begin{eqnarray}
\frac{d\ln\Delta^{ph}_i}{d\ell}&=&2+F_0\left(\hat{\mu}_{\ell}\right)\left(2B^{(1)}_ig_i(\ell)+\frac{1}{4}\sum_{j=1}^{16} B^{(2)}_{ij}g_j(\ell)\right)\\
\frac{d\ln\vec{\Delta}^{ph}_i}{d\ell}&=&2+F_0\left(\hat{\mu}_{\ell}\right)\frac{1}{4}\sum_{j=1}^{16} B^{(2)}_{ij}g_j(\ell)
\end{eqnarray}
where
\begin{eqnarray}
B_i^{(1)}&=&-4+\frac{1}{2}\mbox{Tr}\left[\left(\Gamma_i1\sigma_1\right)^2\right]+\frac{1}{2}\mbox{Tr}\left[\left(\Gamma_i\tau_3\sigma_2\right)^2\right]\\
B_{ij}^{(2)}&=&\mbox{Tr}\left[\left(\Gamma_i\Gamma_j\right)^2\right]-\frac{1}{2}\mbox{Tr}\left(\Gamma_i\Gamma_j1\sigma_1\Gamma_i1\sigma_1\Gamma_j\right)
-\frac{1}{2}\mbox{Tr}\left(\Gamma_i\Gamma_j\tau_3\sigma_2\Gamma_i\tau_3\sigma_2\Gamma_j\right)
\end{eqnarray}
The flow equations for the particle-particle source terms are
\begin{eqnarray}
\frac{d\ln\Delta^{pp}_i}{d\ell}&=&2-\tilde g_i(\ell)\tilde B^{(s)}_i(\ell)\\
\frac{d\ln\vec{\Delta}^{pp}_i}{d\ell}&=&2-\tilde g_i(\ell)\tilde B^{(t)}_i(\ell)
\end{eqnarray}
where
\begin{eqnarray}
\tilde B^{(a)}_i(\ell)=8F_1\left(\hat{\mu}_{\ell}\right)+F_2\left(\hat{\mu}_{\ell}\right)
\left(\mbox{Tr}\left[\left(\Gamma^{(a)}_i1\sigma_1\right)^2\right]-\mbox{Tr}\left[\left(\Gamma^{(a)}_i\tau_3\sigma_2\right)^2\right]\right);\, \; a=s\;\mbox{or}\;t.\nonumber\\
\end{eqnarray}

The correction to free energy, quadratic in the sources $\Delta$, which comes from integrating down to $\ell_c$, i.e. when the upper cutoff has been reduced to $\frac{\Lambda^2}{2m^*}e^{-2\ell_c}$, is
\begin{eqnarray}\label{eq:free energy}
\delta f(\Delta)&=&\frac{m^*}{4\pi}\sum_{j=1}^{16}
B^{(1)}_j\int_0^{\ell_c}d\ell e^{-4\ell}\left(\left|\Delta^{ph}_j(\ell)\right|^2+\left|{\vec{\Delta}}^{ph}_j(\ell)\right|^2\right)F_0\left(\hat{\mu}_{\ell}\right)\nonumber\\
&-&\frac{m^*}{8\pi}\int_0^{\ell_c}d\ell e^{-4\ell}\left(\sum_{j=1}^{10}\left|\Delta^{pp}_j\left(\ell\right)\right|^2\tilde B^{(s)}_j(\ell)+
\sum_{j=11}^{16}\left|\vec{\Delta}^{pp}_j\left(\ell\right)\right|^2\tilde B^{(t)}_j(\ell)\right)
\end{eqnarray}

\section{Flow equations for all $9$ couplings}

The flow equations for the couplings in the $T=0$ limit are given by
\be
\label{eq:flow_full}
\frac{dg_i}{d\ell} = \sum_{j,k = 1}^{16} \sum_{\alpha=0}^2 a_{ijk}^{(\alpha)} F_\alpha \left(\frac{\mu e^{2\ell}}{\Lambda^2 / 2m^*}\right) g_j g_k,
\ee
where the functions $F_\alpha (x)$ are given by Eq.\ \ref{eq: Fs}. The indices $i,j,k$ run over the 9 representations in the order that they are presented in Table I, with multidimensional representations having a corresponding number of identical couplings. The first set of coefficients in Eq.\ \ref{eq:flow_full}, which come from evaluating one-loop particle-hole diagrams, is given by
\begin{equation}\begin{aligned}
a^{(0)}_{ijk} =& \delta_{ij} \delta_{jk} \big\{ \mathrm{Tr}[(\Gamma_i 1\sigma_1)^2]
	+ \mathrm{Tr}[(\Gamma_i \tau_3\sigma_2)^2]  -8 \big\}  \\
&+ \delta_{ij} \big\{ \tfrac{1}{2} \mathrm{Tr} [(\Gamma_i \Gamma_k)^2]
	-\tfrac{1}{4} \mathrm{Tr}
	(\Gamma_i \Gamma_k 1 \sigma_1 \Gamma_i 1 \sigma_1 \Gamma_k)
	-\tfrac{1}{4} \mathrm{Tr}
	(\Gamma_i \Gamma_k \tau_3 \sigma_2 \Gamma_i \tau_3 \sigma_2 \Gamma_k) \big\}
	 \\
&+ \big\{ \tfrac{1}{16} \mathrm{Tr}(\Gamma_i \Gamma_j \Gamma_k)
	\mathrm{Tr} (\Gamma_i \Gamma_k \Gamma_j)
	- \tfrac{1}{32} \mathrm{Tr}(\Gamma_i \Gamma_j 1\sigma_1 \Gamma_k)
	\mathrm{Tr} (\Gamma_i \Gamma_k 1\sigma_1 \Gamma_j) \\
& \quad - \tfrac{1}{32} \mathrm{Tr}(\Gamma_i \Gamma_j \tau_3 \sigma_2 \Gamma_k)
	\mathrm{Tr} (\Gamma_i \Gamma_k \tau_3 \sigma_2 \Gamma_j) \big\}.
\end{aligned}\end{equation}
The remaining coefficients, which come from evaluating the one-loop particle-particle diagram, are given by
\begin{align}
a^{(1)}_{ijk} &= -\tfrac{1}{16} [\mathrm{Tr}(\Gamma_i \Gamma_j \Gamma_k)]^2, \\
a^{(2)}_{ijk} &= -\tfrac{1}{32} \big\{ [\mathrm{Tr}(\Gamma_i \Gamma_j 1\sigma_1 \Gamma_k)]^2
	+ [\mathrm{Tr}(\Gamma_i \Gamma_j \tau_3 \sigma_2 \Gamma_k)]^2 \big\}.
\end{align}

\end{widetext}
\end{appendix}
\end{document}